\documentclass[%
 aip,
 amsmath,amssymb,
 reprint,%
]{revtex4-1}

\usepackage{graphicx}
\usepackage{dcolumn}
\usepackage{bm}

\usepackage[utf8]{inputenc}
\usepackage[T1]{fontenc}
\usepackage{mathptmx}
\usepackage{etoolbox}

\usepackage{graphicx}
\usepackage{epsfig}
\usepackage{amsfonts}
\usepackage{amsmath}
\usepackage{multirow}
\usepackage{bm}
\usepackage{ulem}
\usepackage{color}
\usepackage{hyperref}
\usepackage{mathtools}
\usepackage{cases}
\usepackage{physics}
\usepackage{amssymb}
\usepackage{longtable}
\usepackage{booktabs}

\usepackage{multibib}
\newcites{main}{References}           
\newcites{sup}{References (Supplemental Information)} 

\newcommand{\ie}{{\it i.e.}, }

\newcommand{\exciting}{{\usefont{T1}{lmtt}{b}{n}exciting}~}

\def\GW{{$G_0W_0$}}
\usepackage{xcolor}

\makeatletter
\def\@email#1#2{%
 \endgroup
 \patchcmd{\titleblock@produce}
  {\frontmatter@RRAPformat}
  {\frontmatter@RRAPformat{\produce@RRAP{*#1\href{mailto:#2}{#2}}}\frontmatter@RRAPformat}
  {}{}
}%
\makeatother
\begin{document}

\preprint{AIP/123-QED}

\title{Impact of electron-phonon interaction on the electronic structure of interfaces between organic molecules and a MoS$_2$ monolayer}

\author{Ignacio Gonzalez Oliva}
\affiliation{Physics Department and CSMB, Humboldt-Universit\"{a}t zu Berlin, zum Großen Windkanal 2, 12489 Berlin, Germany}

\author{Sebastian Tillack}
\affiliation{Physics Department and CSMB, Humboldt-Universit\"{a}t zu Berlin, zum Großen Windkanal 2, 12489 Berlin, Germany}

\author{Fabio Caruso}
\affiliation{Physics Department and CSMB, Humboldt-Universit\"{a}t zu Berlin, zum Großen Windkanal 2, 12489 Berlin, Germany}
\affiliation{Institut f\"{u}r Theoretische Physik und Astrophysik, Christian-Albrechts-Universit\"{a}t zu Kiel, D-24098 Kiel, Germany.}

\author{Pasquale Pavone}
\affiliation{Physics Department and CSMB, Humboldt-Universit\"{a}t zu Berlin, zum Großen Windkanal 2, 12489 Berlin, Germany}

\author{Claudia Draxl}
\affiliation{Physics Department and CSMB, Humboldt-Universit\"{a}t zu Berlin, zum Großen Windkanal 2, 12489 Berlin, Germany}
\affiliation{European Theoretical Spectroscopic Facility (ETSF)}

\date{\today}

\begin{abstract}
By means of first-principles calculations, we investigate the role of electron-phonon interaction in the electronic structure of hybrid interfaces, formed by MoS$_2$ and monolayers of the organic molecules pyrene and pyridine, respectively.
Quasiparticle energies are initially obtained within the \GW~approximation and subsequently used to evaluate the electron-phonon self-energy and momentum-resolved spectral functions to assess the temperature renormalization of the band structure. We find that the band-gap renormalization by zero-point vibrations of both hybrid systems is comparable to that of pristine MoS$_2$, with a value of approximately 80 meV. Pronounced features of molecular origin emerge in the spectral function of the valence region, which we attribute to satellites arising from out-of-plane vibrational modes of the organic monolayers. For pyrene, this satellite exhibits a predominantly molecular character, while for pyridine, it has a hybrid nature, originating from the coupling of molecular vibrations to the MoS$_2$ valence band.
\end{abstract}

\maketitle

\section{Introduction}
\label{section:introduction}

Hybrid inorganic-organic systems (HIOS) are a versatile class of materials that combine components of different nature~\citemain{Sanchez2001,GomezRomero2001}. Ideally, when appropriately designed, these systems can leverage complementary properties, for example, combining the thermal stability of inorganic frameworks with the structural flexibility of organic molecules~\citemain{Sanchez2011,Kickelbick2003,Benson2024}. A particularly promising subset of HIOS is realized when two-dimensional transition metal dichalcogenides (TMDCs) are interfaced with organic molecules to form weakly bound van der Waals heterostructures~\citemain{Xu2021,Ji2022,Obaidulla2024,Cui2025}. Recent studies have reported on a variety of optical phenomena in these systems, including the formation of hybrid excitons via Förster energy transfer~\citemain{Park2021,Bennecke2025} or charge transfer~\citemain{Mutz2020,GonzalezOliva2022,Raoufi2023}, as well as interface-induced photoluminescence modulation~\citemain{Obaidulla2020}. These effects demonstrate the potential for tailoring light-matter interactions in hybrid materials, making use of the interplay between many-body interactions that control their electronic and excitonic states~\citemain{Hofmann2021,GonzalezOliva2024,Liu2025}.

Theoretical investigations based on many-body perturbation theory have established the crucial role of electron-electron and electron-hole interactions in organic/TMDC heterostructures, for describing the energy-level alignment~\citemain{Adeniran2021,Cheng2021,GonzalezOliva2022, Schebek2025_PRB} and the lowest-energy excitons~\citemain{Palummo2019,Wang2020,GonzalezOliva2022,Bennecke2025,Chowdhury2024, Schebek2025_PRB}, respectively. However, the role of electron-phonon coupling (EPC) in such materials remains largely unexplored.   

EPC plays a central role in carrier mobilities~\citemain{Gunst2016}, ultrafast relaxation~\citemain{Nayyar2022,Pan2025}, exciton linewidths~\citemain{Moody2015}, and polaronic spectral features~\citemain{Caruso2021B}. The development of first-principles approaches for EPC~\citemain{Eiguren2008,Eiguren2009,Gonze2011,Ponce2014,Ponce2015,Marini2015,Giustino2017,Schebek2025} has facilitated the calculation of temperature-dependent band structures, quasiparticle (QP) lifetimes, and momentum-resolved spectral functions that can be compared with angle-resolved photoemission spectroscopy (ARPES)~\citemain{Park2009} and ultrafast experimental probes~\citemain{Emeis2025}. In the case of MoS$_2$ monolayers, for example, electrons and phonons interact to regulate ultrafast carrier dynamics on femtosecond timescales~\citemain{Britt2022, Caruso2021} and to determine exciton lifetimes~\citemain{Chan2023}. On the molecular side, EPC leads to a temperature-dependent renormalization of the electronic band structure and has been shown to significantly impact carrier mobilities in organic crystals~\citemain{Brown2020}.

Given the impact of EPC on the individual components, it is natural to consider its role in hybrid materials. Experimentally, temperature-dependent Raman measurements have revealed EPC signatures in organic/TMDC heterostructures~\citemain{Sarkar2017}. On the theory side, effective electron-phonon models have been employed to study the influence of vibrations on the electronic states at metal-organic interfaces~\citemain{Eschmann2021,Eschmann2021_B}, while first-principles calculations are still lacking. In order to achieve a comprehensive understanding of organic/TMDC heterostructures, it is essential to incorporate EPC on equal footing with electron-electron and electron-hole interactions, since their interplay collectively determines the optoelectronic properties of these materials.

In this work, we present a first-principles study of EPC of two prototypical hybrid interfaces composed of a MoS$_2$ monolayer and the $\pi$-conjugated organic molecules pyrene (C$_{16}$H$_{10}$) and pyridine (C$_{5}$H$_{5}$N), respectively. We calculate the electron-phonon self-energy and momentum-resolved spectral functions using QP energies obtained from prior \GW~calculations~\citemain{GonzalezOliva2022}. Our findings reveal the emergence of distinct spectral features, demonstrating the necessity for high-level theory to effectively capture these many-body effects.
\section{Theoretical background}
\label{section:theoretical_background}
Ground-state calculations are performed within density-functional theory, giving the Kohn-Sham (KS) wave functions $\psi_{n\mathbf{k}}$, where $n$ is the band index and $\mathbf{k}$ is a point in the first Brillouin-zone (BZ). The electronic QP energies $\varepsilon^{\text{QP}}_{n \mathbf{k}}$ are obtained by $G_0W_0$ calculations. The potential response and the dynamical matrices are computed by means of density-functional perturbation theory (DFPT) and the finite-displacement method~\citemain{Baroni2001}. 

For obtaining the electron-phonon self-energy, 
$\Sigma^{\text{eph}}_{n\mathbf{k}} (\omega, T)$, we follow the approach described in Refs.~\onlinecite{Ponce2015,Marini2015,Giustino2017} using Hartree atomic units throughout. It consists of the Fan-Migdal (FM) term and the Debye-Waller (DW) term,
\begin{align}
    \Sigma^{\text{eph}}_{n\mathbf{k}} (\omega, T) = \Sigma^{\text{FM}}_{n\mathbf{k}} (\omega, T) + \Sigma^{\text{DW}}_{n\mathbf{k}} (T)  \:.
    \label{eq:eph_self_energy}
\end{align}
The FM term is written as
\begin{align}
    \Sigma^{\text{FM}}_{n\mathbf{k}} (\omega,T) & = \sum_{m\nu} \int_{\text{BZ}} \frac{d\mathbf{q}}{\Omega_{\text{BZ}}}  |g_{mn\nu} (\mathbf{k},\mathbf{q})|^2 \nonumber \\[5pt]
    & \times \left[ \frac{n_{\mathbf{q}\nu} (T) + f_{m\mathbf{k}+\mathbf{q}}(T)}{\omega - ( \varepsilon^{\text{QP}}_{m\mathbf{k}+\mathbf{q}}- \varepsilon_{\text{F}}) + \omega_{\mathbf{q}\nu} + i\eta} \right. \nonumber \\[5pt]
     & \left. + \frac{n_{\mathbf{q}\nu}(T) + 1 - f_{m\mathbf{k}+\mathbf{q}}(T)}{\omega - ( \varepsilon^{\text{QP}}_{m\mathbf{k}+\mathbf{q}}- \varepsilon_{\text{F}}) - \omega_{\mathbf{q}\nu} + i\eta }  \right] \:,
    \label{eq:eph_self_energy_fm}
\end{align}
where $\nu$ specifies the phonon modes, $\omega_{\mathbf{q}\nu}$ are the phonon energies, and $\varepsilon_{\text{F}}$ is the Fermi energy. The electron-phonon matrix elements $g_{mn\nu} (\mathbf{k}, \mathbf{q})$,
\begin{align}
    g_{mn\nu} (\mathbf{k}, \mathbf{q}) = \langle \psi_{m\mathbf{k+q}} | \: V_{\mathbf{q} \nu}^{(1)} \: |  \psi_{n\mathbf{k}} \rangle \:,
    \label{eq:eph_matrix_elements}
\end{align}
can be understood as the probability amplitude for the scattering of an electron in state $\psi_{n\mathbf{k}}$ into the state $\psi_{m\mathbf{k+q}}$ due to a perturbation of the KS potential $V_{\mathbf{q} \nu}^{(1)}$ caused by a phonon in the mode $\nu$ with crystal momentum $\mathbf{q}$. The temperature dependence of the FM self-energy arises from the Fermi-Dirac occupation factors for the electrons, $f_{m\mathbf{k}+\mathbf{q}}$, and the Bose-Einstein distribution functions for the phonons, $n_{\mathbf{q}\nu}$.

The static DW self-energy
\begin{align}
   \Sigma^{\text{DW}}_{n\mathbf{k}} (T) = - \sum_{m \neq n ,\nu} \int_{\text{BZ}} \frac{d\mathbf{q}}{\Omega_{\text{BZ}}}  \frac{2n_{\nu\mathbf{q}}(T) + 1}{\varepsilon^{\text{QP}}_{n\mathbf{k}} - \varepsilon^{\text{QP}}_{m\mathbf{k}}} g^{2,\text{DW}}_{mn\nu} (\mathbf{k},\mathbf{q}) \:,
    \label{eq:eph_self_energy_dw}
\end{align}
contains the effective DW matrix elements, $g^{2,\text{DW}}_{mn\nu} (\mathbf{k}, \mathbf{q})$, which are calculated within the rigid-ion approximation. In this formulation, the DW contribution is constructed using $V_{\mathbf{q} \nu}^{(1)}$, rather than the second-order response of the potential $V_{\mathbf{q}\nu,~\mathbf{q'} \nu'}^{(2)}$~\citemain{Allen1976,Gonze2011,Ponce2014}.

The spectral function is given by
\begin{align}
    A_{n \mathbf{k}}(\omega) = -\frac{1}{\pi} \frac{\text{Im} \Sigma^{\text{eph}}_{n \mathbf{k}} (\omega)}{\Big[\omega - \varepsilon^{\text{QP}}_{n\mathbf{k}} - \text{Re} \Sigma^{\text{eph}}_{n\mathbf{k}} (\omega)\Big]^2 + \Big[\text{Im} \Sigma^{\text{eph}}_{n\mathbf{k}}(\omega)\Big]^2} \:. \label{eq:spectral_function}
\end{align}
Within the sudden approximation and neglecting photoemission matrix elements, the spectral function provides a theoretical analogue to ARPES measurements.
\section{Computational details}
\label{section:computational_details}
All calculations are carried out within the all-electron full-potential (linearized) augmented planewave plus local orbital (LAPW+LO) method, implemented in the \exciting~\citemain{Gulans2014} code. The geometries of the MoS$_2$ monolayer and the hybrid materials, as well as their computational parameters of the ground-state and \GW~calculations, are as reported in Ref.~\onlinecite{GonzalezOliva2022}. In short, in all ground-state calculations, we use the generalized gradient approximation in the Perdew-Burke-Ernzerhof (PBE) parametrization~\citemain{Perdew1996} as the exchange correlation functional. The electronic eigenvalues and wavefunctions are calculated by sampling the BZ with a 15$\times$15$\times$1 Monkhorst-Pack \textbf{k}-point grid for the pristine MoS$_2$ and 3$\times$3$\times$1 for the heterostructures. Spin-orbit coupling is neglected in our calculations. Phonons are calculated within DFPT, using a 5$\times$5$\times$1 \textbf{q}-point grid for pristine MoS$_2$. For the hybrid systems, we compute only $\Gamma$-point phonons using the finite-displacement method. Maximally localized Wannier functions are obtained from the PBE wave functions via optimized projection functions and self-projections~\citemain{Tillack2020,Tillack2025}, used for Wannier-Fourier interpolation of the electron-phonon matrix elements~\citemain{Giustino2007}. For the calculation of the electron-phonon self-energies, the electron-phonon matrix elements are interpolated to a 59$\times$59$\times$1 \textbf{q}-grid for MoS$_2$ monolayer and a 36$\times$36$\times$1 \textbf{q}-grid for the hybrid systems. BZ integrals in Eqs.~\eqref{eq:eph_self_energy_fm} and \eqref{eq:eph_self_energy_dw} are evaluated using tetrahedron integration, which allows for the calculation of the ${\eta \rightarrow 0^+}$ limit and ensures quick convergence. 
\section{Results and discussion}
\subsection{MoS$_2$ monolayer}
\label{section:mos2}
\begin{figure}[t!]
\begin{center}
\includegraphics[width=\columnwidth]{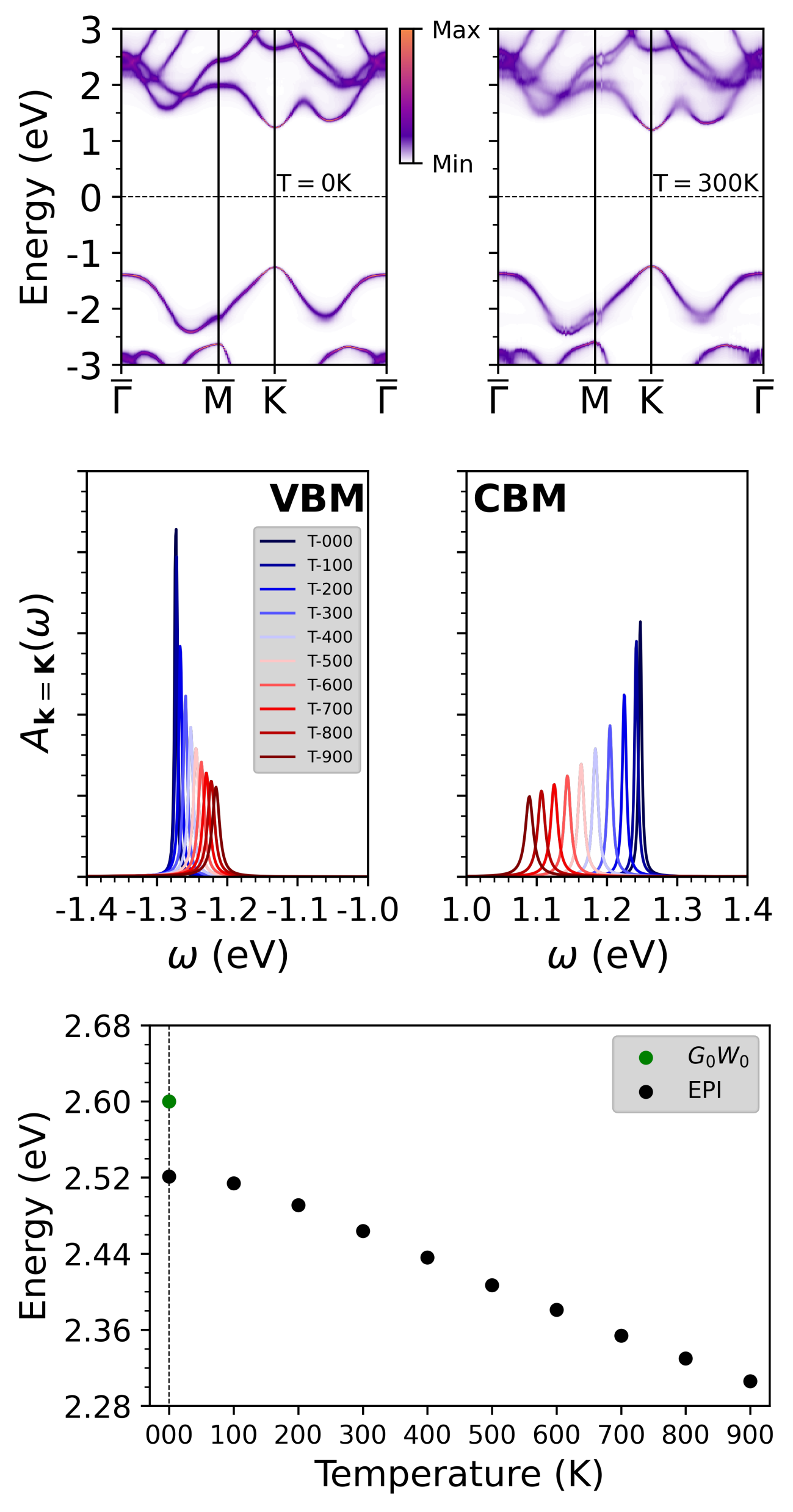}
\caption{Top: Temperature-dependent band structure of the MoS$_2$ monolayer for 0K (left) and 300K (right). The color scale bar, normalized to the global maximum across both temperatures, ranges from the minimum (white) to the maximum value (orange) of the spectral function. Middle: Spectral functions for the VBM and CBM at the K point, for temperatures up to 900K. Bottom: Temperature-dependence of the band gap. The QP gap obtained with the \GW~approximation is indicated by the green dot.
\label{fig:mos2_eph}}
\end{center}
\end{figure}

We begin our discussion by showing the temperature-dependent band structure of the pristine MoS$_2$ monolayer, which is plotted along the high-symmetry points of the unit-cell BZ for 0K (left) and 300K (right) in the top panel of Fig.~\ref{fig:mos2_eph}. We immediately notice that the band gap decreases as the temperature increases. This renormalization of the bands has been found for many semiconductors~\citemain{Giustino2010,Milot2015,Karsai2018}, and has also been reported for MoS$_2$~\citemain{MolinaSanchez2016}. 

To further investigate the renormalization of the direct band gap at K due to temperature effects, the spectral functions of the valence-band maximum (VBM) and conduction-band minimum (CBM) are plotted in the middle panel of Fig.~\ref{fig:mos2_eph}, for temperatures ranging from 0K to 900K, in 100K steps. As expected, renormalization in energy with a slight increase in broadening as the temperature increases are observed. This indicates that the electron lifetimes at the band edges are reduced upon increasing temperature. For the given temperature range, the peak position changes from 1.25 eV ($-1.27$ eV) to 1.09 eV ($-1.21$ eV) for the CBM (VBM), a total renormalization of $-0.16$ eV ($+0.06$ eV). This indicates that the CBM experiences a larger renormalization than the VBM.

In the bottom panel of Fig.~\ref{fig:mos2_eph}, the direct band gap at K as a function of temperature is shown, as derived from the spectral functions of the band edges. The band gap decreases from 2.52 eV at 0K to 2.31 eV at 900K. The room temperature band gap (300K) has a value of 2.46 eV, which compares well with the experimental gap of 2.50 eV measured by photocurrent spectroscopy~\citemain{Klots2014}. The zero-point renormalization (ZPR) of the band gap is estimated as the difference between the \GW~gaps without (2.60 eV) and with EPC (2.52 eV) at 0K, amounting to 80 meV. The calculated ZPR value of the band gap is in good agreement with the 75 meV reported in Ref.~\onlinecite{MolinaSanchez2016} and the 65 meV reported in Ref.~\onlinecite{Zacharias2020}. It is important to note that our present estimate accounts only for the electron-phonon contribution at fixed lattice parameters, \ie does not include an additional shift arising from zero-point lattice expansion~\citemain{Brousseau2022}.

\subsection{Pyrene@MoS$_2$}
\label{section:pyrene_mos2}

\begin{figure}[t!]
\begin{center}
\includegraphics[width=\columnwidth]{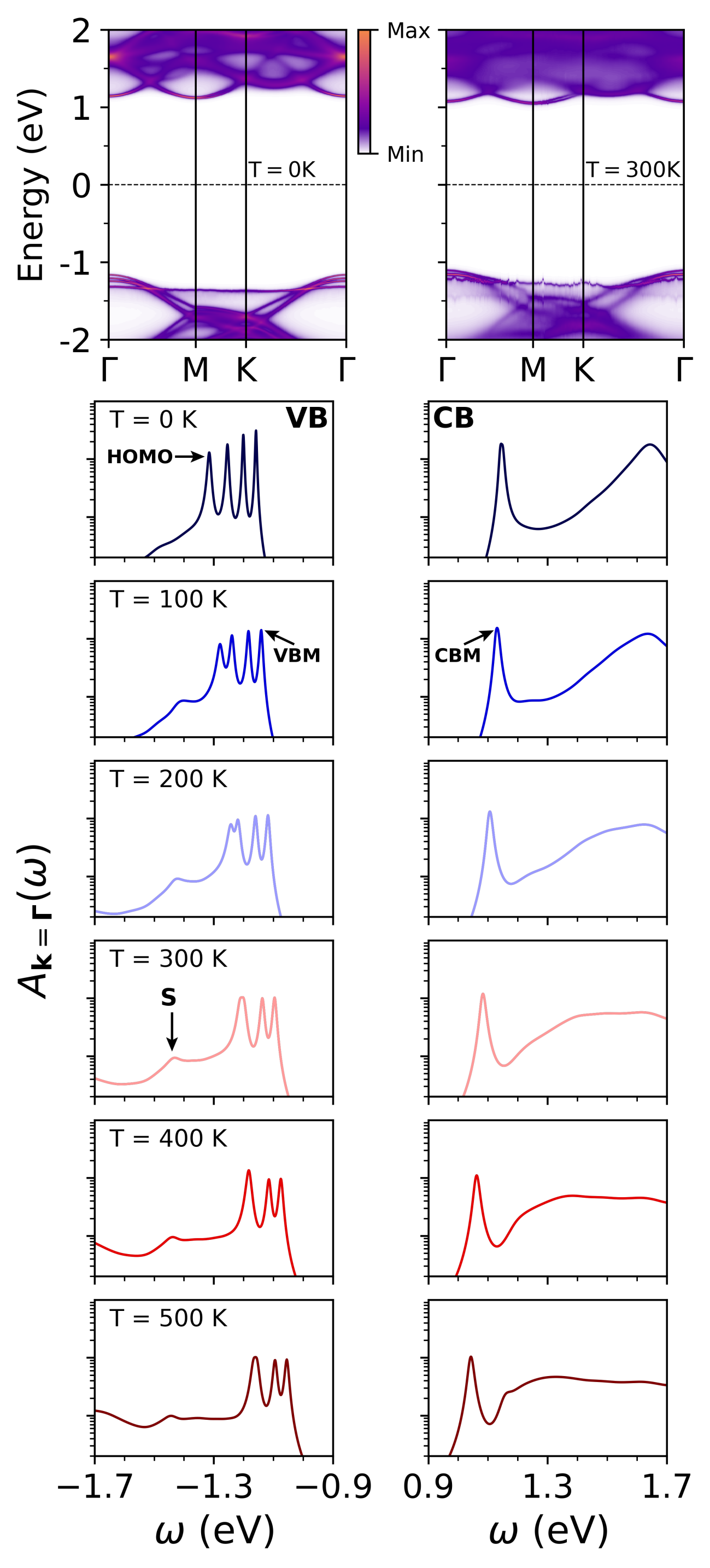}
\caption{Top: Temperature-dependent band structure of pyrene@MoS$_2$ for 0K (left) and 300K (right). The color scale bar, normalized to the global maximum across both temperatures, ranges from the minimum (white) to the maximum (orange) value of the spectral function. Bottom: Spectral functions for valence and conduction bands at the $\Gamma$ point, for temperatures up to 500K.
\label{fig:pyrene_mos2_eph}}
\end{center}
\end{figure}

The top panel of Fig.~\ref{fig:pyrene_mos2_eph} shows the temperature-dependent band structure of pyrene@MoS$_2$ along the high-symmetry points of the supercell BZ for 0K (left) and 300K (right). This material exhibits a direct gap at the $\Gamma$-point, as a consequence of the band folding in the 3$\times$3$\times$1 supercell~\citemain{Valencia2017}. The fundamental \GW~gap, reported in Ref.~\onlinecite{GonzalezOliva2022}, amounts to 2.38 eV. The top valence states at $\Gamma$ correspond to the top three valence bands of MoS$_2$, namely VBM, VB-1, and VB-2. Slightly below in energy, the flat band emerges from the highest-occupied molecular orbital (HOMO). The CBM comes from a degenerate band of MoS$_2$. As temperature increases, we observe not only a reduction of the band gap but also the appearance of a spectral feature at an energy of $-1.44$~eV. Its weak dispersion resembles that of the HOMO, underpinning the molecular origin. We note that there are no bands at this specific energy in the QP band structure. Consequently, this peak is considered as a \textit{satellite}.

To further investigate the band gap renormalization in this system and the origin of the satellite, the spectral functions of the valence and conduction bands at the $\Gamma$-point within the energy range from $-1.7$~eV to 1.7~eV and in a temperature range from 0K to 500K are plotted in the bottom panels of Fig.~\ref{fig:pyrene_mos2_eph}. The three top-most valence-band features, \ie at 0K, in the energy range from $-1.3$~eV to $-1.1$~eV, correspond to VBM, VB-1, and VB-2. The fourth peak at around $-1.31$~eV, stems from the HOMO of pyrene. Overall, the temperature-dependent renormalization of molecular states is more pronounced than that of the MoS$_2$ states. This is evident as starting at a temperature of 300K, the QP peak of the HOMO merges with the one from VB-2, and it is no longer possible to distinguish them. A well-defined QP peak corresponding to the CBM appears at an energy of 1.1~eV, while the higher-energy peaks cannot be properly distinguished. With increasing temperature, the fundamental gap of pyrene@MoS$_2$ changes from 2.30 eV (0K) to 2.09~eV~(500K). We estimate the ZPR of the band gap as the difference between the \GW~band gap and the gap obtained from the spectral function at 0K to be about 80 meV -- the same value as for pristine MoS$_2$. 

We now focus our attention on the low-intensity satellite (labeled S at 300~K) that appears at $-1.44$~eV, approximately 130 meV below the QP HOMO peak, which is located at $-1.31$~eV at 0K. This energy separation should correspond to a phonon energy of the system~\citemain{Caruso2018B}, allowing us to identify the specific vibrational modes responsible for the satellite formation. We recall that the largest phonon energy of MoS$_2$ is approximately 60 meV~\citemain{MolinaSanchez2011,MolinaSanchez2015}, much too small for explaining the feature. The vibrational energies of pyrene range from 14 meV to 360 meV (see Supplemental Information). In particular, the out-of-plane C-H bend modes span energies from 90 meV to 131 meV, encompassing the estimated satellite energy. Therefore, we classify this spectral signature as a \textit{molecular satellite}. The energy separation between the QP peaks and satellites arising from electron correlations is typically overestimated~\citemain{Guzzo2011}. Here, a more accurate description can be obtained within the cumulant expansion~\citemain{Kas2014}, which has been shown to provide quantitatively improved energies for electronic excitations. The formalism has been extended to treat EPC~\citemain{Verdi2017,Nery2018, Caruso2018}, but its applicability to organic/TMDC heterostructures has yet to be assessed. In any case, the estimated 130 meV lies at the upper bound of the out-of-plane C-H bend modes. Therefore, even if the energy separation were overestimated with our framework, the satellite's origin could still be consistently attributed to these molecular vibrations.  

\subsection{Pyridine@MoS$_2$}
\label{section:pyridine_mos2}

\begin{figure}[th!]
\begin{center}
\includegraphics[width=\columnwidth]{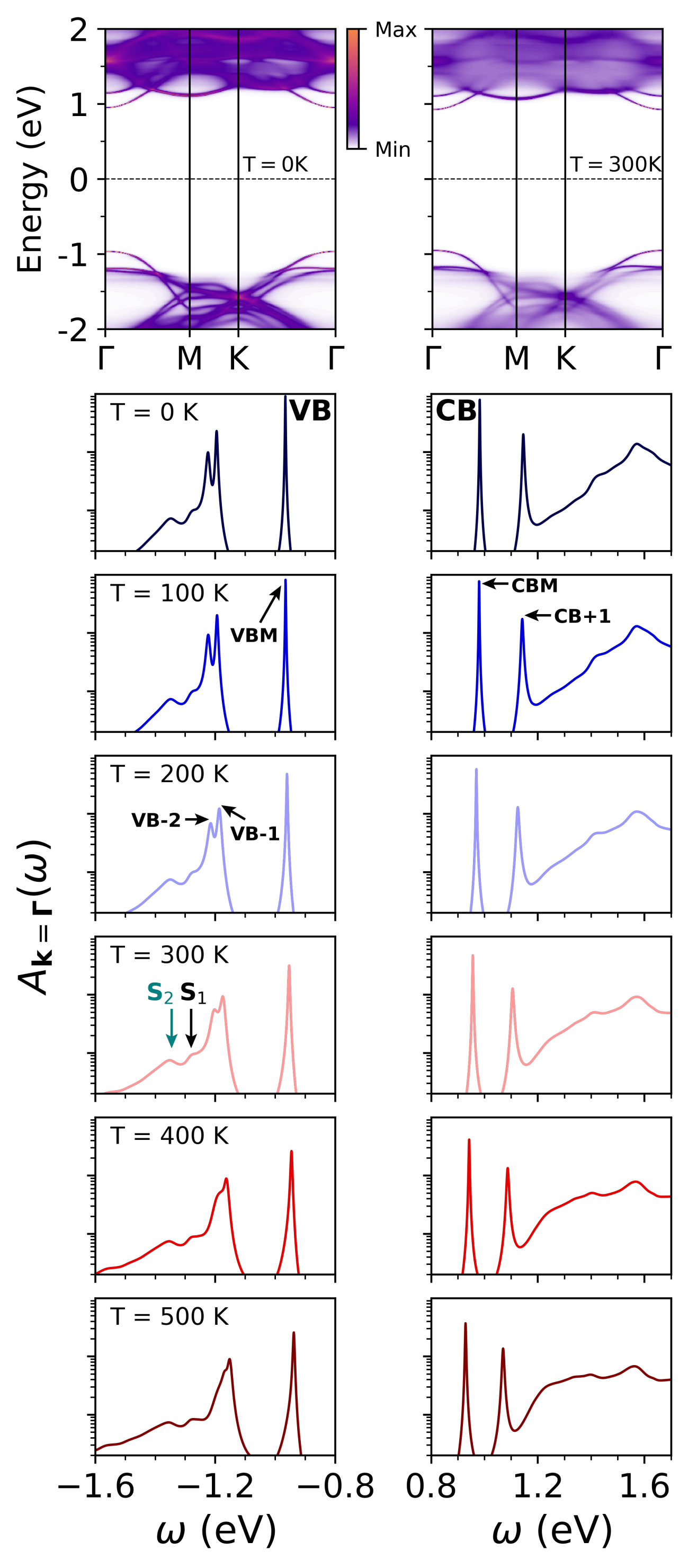}
\caption{Top: Temperature-dependent band structure of pyridine@MoS$_2$ for 0K (left) and 300K (right). The color scale bar, normalized to the global maximum across both temperatures, ranges from the minimum (white) to the maximum (orange) value of the spectral function. Bottom: Spectral functions for valence and conduction bands at the $\Gamma$ point, for temperatures up to 500K.
\label{fig:pyridine_mos2_eph}}
\end{center}
\end{figure}

The top panel of Fig.~\ref{fig:pyridine_mos2_eph} shows the temperature-dependent band structure of pyridine@MoS$_2$ for 0K (left) and 300K (right). Like pyrene@MoS$_2$, this material exhibits a direct gap at the $\Gamma$-point, as a consequence of the band folding. The fundamental gap of pyridine@MoS$_2$, calculated with the \GW~approximation, amounts to 2.00~eV~\citemain{GonzalezOliva2022}. The top valence bands at $\Gamma$ correspond to VBM, VB-1, and VB-2, while the bottom conduction bands emerge from MoS$_2$ states. Unlike pyrene@MoS$_2$, the VB-1 and VB-2 have a more hybridized character with contributions from pyridine~\citemain{GonzalezOliva2022}. As temperature increases, we observe not only a reduction of the band gap but also the appearance of spectral features close to VB-2. 

The bottom panel of Fig.~\ref{fig:pyridine_mos2_eph} shows the spectral functions of the valence and conduction bands at the $\Gamma$-point within the energy range from $-1.6$ eV to 1.6 eV for temperatures between 0K and 500K. In the valence region, the topmost QP peak at an energy of $-0.97$~eV (0K) is assigned to the VBM, while the two peaks at energies of $-1.20$ eV and $-1.23$ eV, are assigned to the VB-1, and VB-2, respectively. In the conduction region, we observe two distinct QP peaks, one at 0.96 eV, corresponding to the CBM, and a second peak at 1.10 eV which is attributed to CB+1. Similar to the results for pyrene@MoS$_2$, higher-energy peaks can no longer be properly distinguished. The fundamental gap changes from 1.92 eV (0K) to 1.84 eV~(500K). Using the spectral functions of the band edges, we estimate the ZPR of the band gap to be about 80 meV, the same value found for pristine MoS$_2$ and pyrene@MoS$_2$.  

We focus our attention on two low-intensity features at energies of $-1.28$~eV and $-1.34$ eV, which we label S$_1$ and S$_2$ at 300~K, respectively. As in the previous case, there are no bands at these specific energies in the QP band structure, so we regard them as satellites. There are different possible explanations for their origin: The energy difference between the VB-2 and S$_1$ peaks is 50 meV, which would be consistent with the hypothesis of a MoS$_2$ origin. However, this is unlikely due to the absence of a comparable feature observed in the results of the MoS$_2$ monolayer. The vibrational energies of pyridine have a range from 16 meV to 386 meV (see Supplemental Information). An analysis of the band-resolved spectral functions reveals that S$_1$ originates from VB-1, whereas S$_2$ is associated with VB-2. The energy difference between S$_1$ and VB-1 amounts to 80 meV. Accordingly, the responsible vibrations would be the out-of-plane ring bend modes. For S$_2$, the difference to VB-2 is 110 meV. The modes with such energies are the out-of-plane C-H. These findings indicate that both S$_1$ and S$_2$ are \textit{hybrid satellites}, arising from the coupling of pyridine vibrational modes with MoS$_2$ hybridized valence-band states.

\section{Summary and Conclusions}
\label{section:conclusions}

In summary, we have investigated the impact of electron-phonon interaction on the electronic structure of MoS$_2$ and its weakly-bound hybrid interfaces with pyrene and pyridine. Our results reproduce the temperature-dependent band-gap renormalization of MoS$_2$ in agreement with prior reports. For the hybrid materials, we find that this effect is essentially unaffected by the presence of the organic layers. For pyrene@MoS$_2$, the renormalization of the HOMO-derived band is more pronounced than that of the MoS$_2$ bands, however, there is no change of the energy-level alignment at the interface. In both heterostructures, we identify satellites in the spectral function, arising from out-of-plane vibrational modes. While in pyrene@MoS$_2$, these satellites have essentially pure molecular character, in pyridine@MoS$_2$, they are of hybrid nature. These observations suggest the presence of subtle yet distinct polaronic fingerprints at such interfaces, which could be investigated with high-resolution ARPES.

\section*{Acknowledgements}
This work was supported by Deutsche Forschungsgemeinschaft (DFG) within the Collaborative Research Center HIOS (SFB 951), project 182087777. I.G.O. acknowledges financial support from the German Academic Exchange Service (DAAD). Partial support was also provided by the European Union’s Horizon 2020 Research and Innovation Program under the grant agreement no. 951786 (NOMAD CoE). Computing time on the supercomputers Lise and Emmy at NHR@ZIB and NHR@Göttingen and the Paderborn Center for Parallel Computing (PC2) are appreciated. We are grateful for the support of Manoar Hossain for assistance regarding the PC2 computer cluster. We also thank M. Schebek for critically reading the manuscript.

\section*{Data availability}
Input and output files of all calculations can be downloaded free of charge from the NOMAD data infrastructure~\citemain{Draxl2019} by the following link: \url{https://dx.doi.org/10.17172/NOMAD/2025.09.30-1}.\\



\nocitemain{*}
\bibliographystylemain{apsrev4-1}
\bibliographymain{bib}

\setcounter{section}{0}
\renewcommand{\thesection}{S\arabic{section}}

\renewcommand{\thefigure}{S\arabic{figure}}
\setcounter{figure}{0}

\renewcommand{\thetable}{S\arabic{table}}
\setcounter{table}{0}

\renewcommand{\theequation}{S\arabic{equation}}
\setcounter{equation}{0}

\newpage
\clearpage
\onecolumngrid            
\setcounter{page}{1}      
\renewcommand{\thepage}{S\arabic{page}} 
{\Large \textbf{Supplemental Information}}

\section{Vibrational energies of pyrene and pyridine}

Pyrene has 78 modes, but due to translational and rotational symmetries, 6 modes are zero (or numerically close to zero), meaning that it has 72 normal modes~\citesup{Baba2009}. In Table~\ref{table:pyrene_vibrations}, we show the vibrational energies of pyrene computed in this work, and the values reported in Ref.~\onlinecite{Baba2009}. Moreover, we follow their classification of the vibrations, there are 8 vibration types, in-plane, and out-of-plane (oop) C-H bend , C-H stretch, C-C stretch, in-plane, and oop ring deform, ring breathing, oop ring torsion, oop ring wag and butterfly. Overall, the vibrational energies are in good agreement, and the small differences could be related to the different methodologies used. 

\renewcommand{\arraystretch}{0.85} 
\begin{longtable}{c c c c c}
\caption[Vibrational energies of pyrene]{Vibrational energies of pyrene calculated with \exciting~and compared with theoretical and experimental results from Ref.~\onlinecite{Baba2009}.}
\label{table:pyrene_vibrations} \\
\toprule
Number & Vibration type & This work [meV] & Theory [meV] & Exp. [meV] \\
\midrule
\endfirsthead

\toprule
Number & Vibration type & This work [meV] & Theory [meV] & Exp. [meV] \\
\midrule
\endhead
1 & butterfly & 13.94 & 11.904 & -- \\ \midrule
2 & oop ring deform & 23.43 & 18.476 & -- \\ \midrule
3 & oop ring deform & 32.82 & 25.916 & -- \\ \midrule
4 & oop ring torsion & 33.71 & 30.13 & -- \\ \midrule
5 & oop ring wag & 35.55 & 31.99 & -- \\ \midrule
6 & ring deform & 38.60 & 43.89 & -- \\ \midrule
7 & oop ring deform & 42.27 & 48.98 & -- \\ \midrule
8 & ring deform & 49.26 & 50.34 & 49.84 \\ \midrule
9 & ring deform & 49.56 & 56.54 & 55.92 \\ \midrule
10 & oop ring deform & 52.58 & 61.13 & -- \\ \midrule
11 & oop ring deform & 55.30 & 62.00 & -- \\ \midrule
12 & ring deform & 58.18 & 62.00 & -- \\ \midrule
13 & ring deform & 58.51 & 62.12 & 62.00 \\ \midrule
14 & oop ring deform & 59.65 & 65.72 & -- \\ \midrule
15 & ring deform & 60.35 & 67.82 & -- \\ \midrule
16 & oop ring deform & 67.16 & 70.80 & -- \\ \midrule
17 & ring breathing & 67.33 & 72.78 & 73.53 \\ \midrule
18 & oop C-H bend & 74.19 & 83.70 & -- \\ \midrule
19 & ring deform & 78.55 & 86.05 & -- \\ \midrule
20 & oop ring deform & 81.96 & 88.04 & -- \\ \midrule
21 & ring deform & 83.37 & 91.63 & 91.14 \\ \midrule
22 & oop C-H bend & 90.16 & 92.50 & -- \\ \midrule
23 & oop C-H bend & 91.53 & 94.116 & -- \\ \midrule
24 & oop C-H bend & 91.90 & 99.57 & -- \\ \midrule
25 & ring deform & 96.90 & 99.69 & 99.82 \\ \midrule
26 & oop C-H + C-C bend & 102.06 & 99.94 & -- \\ \midrule
27 & ring deform & 106.37 & 102.05 & -- \\ \midrule
28 & oop C-H bend & 109.56 & 104.65 & -- \\ \midrule
29 & oop C-H bend & 113.70 & 110.48 & -- \\ \midrule
30 & oop C-H bend & 115.25 & 112.22 & -- \\ \midrule
31 & oop C-H bend & 119.66 & 119.41 & -- \\ \midrule
32 & C-H bend + ring deform & 121.13 & 119.90 & -- \\ \midrule
33 & oop C-H bend & 123.36 & 120.03 & -- \\ \midrule
34 & oop C-H bend & 126.81 & 120.15 & -- \\ \midrule
35 & oop C-H bend & 131.29 & 121.272 & -- \\ \midrule
36 & ring deform & 131.90 & 124.12 & -- \\ \midrule
37 & C-H bend + ring deform & 136.49 & 133.42 & 132.92 \\ \midrule
38 & ring deform & 137.53 & 136.27 & -- \\ \midrule
39 & C-H bend + ring deform & 141.41 & 137.88 & 137.76 \\ \midrule
40 & C-H bend & 145.18 & 142.60 & -- \\ \midrule
41 & C-H bend & 147.84 & 142.84 & 141.35 \\ \midrule
42 & C-H bend & 149.02 & 146.69 & 150.28 \\ \midrule
43 & C-H bend + ring deform & 149.24 & 147.43 & -- \\ \midrule
44 & C-H bend + ring deform & 159.77 & 150.66 & -- \\ \midrule
45 & C-H bend + ring deform & 161.48 & 154.38 & 153.51 \\ \midrule
46 & C-H bend + ring deform & 161.67 & 154.62 & 153.51 \\ \midrule
47 & C-H bend + ring deform & 167.67 & 155.12 & -- \\ \midrule
48 & C-H bend + C-C stretch & 169.29 & 164.30 & -- \\ \midrule
49 & ring deform & 174.68 & 165.29 & 164.42 \\ \midrule
50 & C-H bend + ring deform & 174.94 & 171.24 & 170.12 \\ \midrule
51 & C-C stretch & 175.56 & 174.34 & 174.84 \\ \midrule
52 & C-C stretch & 176.34 & 175.46 & -- \\ \midrule
53 & C-H bend + C-C stretch & 197.24 & 178.31 & -- \\ \midrule
54 & C-H bend + ring deform & 198.03 & 178.43 & -- \\ \midrule
55 & C-H bend + C-C stretch & 200.87 & 181.41 & -- \\ \midrule
56 & C-H bend + C-C stretch & 201.68 & 185.13 & -- \\ \midrule
57 & C-H bend + C-C stretch & 204.46 & 187.61 & 187.61 \\ \midrule
58 & C-H bend + C-C stretch & 205.64 & 194.92 & -- \\ \midrule
59 & C-C stretch & 208.92 & 198.77 & 194.80 \\ \midrule
60 & C-C stretch & 209.45 & 199.88 & -- \\ \midrule
61 & C-H bend + C-C stretch & 209.80 & 201.00 & -- \\ \midrule
62 & C-C stretch & 210.25 & 204.10 & 201.37 \\ \midrule
63 & C-H stretch & 347.64 & 386.13 & -- \\ \midrule
64 & C-H stretch & 348.66 & 386.26 & -- \\ \midrule
65 & C-H stretch & 349.80 & 386.63 & -- \\ \midrule
66 & C-H stretch & 351.16 & 386.63 & -- \\ \midrule
67 & C-H stretch & 353.59 & 387.50 & -- \\ \midrule
68 & C-H stretch & 354.22 & 387.62 & -- \\ \midrule
69 & C-H stretch & 354.80 & 388.61 & -- \\ \midrule
70 & C-H stretch & 355.86 & 388.61 & -- \\ \midrule
71 & C-H stretch & 356.83 & 389.60 & -- \\ \midrule
72 & C-H stretch & 358.95 & 389.73 & -- \\
\bottomrule
\end{longtable}

\newpage

Pyridine has 33 modes, due to translational and rotational symmetries 6 modes are zero, meaning that it has 27 normal modes~\citesup{Urena2003}. In Table~\ref{table:pyridine_vibrations}, we show the vibrational energies of pyridine computed in this work, and the values reported in Ref.~\onlinecite{Urena2003}. We follow their classification of the vibrations, there are 5 vibrational types, out-of-plane (oop) ring bend, ring stretch, oop C-H wag, C-H bend, and C-H stretch. Similar to the case of pyrene, the differences in the vibrational energies could be related to the different methodologies applied. 

\renewcommand{\arraystretch}{0.85} 
\begin{longtable}{c c c c c}
\caption[Vibrational energies of pyridine]{Vibrational energies of pyridine calculated with \exciting~and compared with theoretical and experimental values from Ref.~\onlinecite{Urena2003}.}
\label{table:pyridine_vibrations} \\
\toprule
Number & Vibration type & This work [meV] & Theory [meV] & Exp. [meV] \\
\midrule
\endfirsthead

\toprule
Number & Vibration type & This work [meV] & Theory [meV] & Exp. [meV] \\
\midrule
\endhead
1  & oop ring bend & 16.03  & 53.93 & 46.37 \\ \midrule
2  & oop ring bend & 34.99  & 57.16 & 50.34 \\ \midrule
3  & ring stretch & 63.70  & 81.67 & 74.76  \\ \midrule
4  & oop ring bend & 85.31  & 88.96 & 80.96  \\ \midrule
5  & oop ring bend + oop C-H wag & 95.09  & 96.23 & 86.91 \\ \midrule
6  & oop ring bend + oop C-H wag  & 98.50  & 103.72 & 92.74 \\ \midrule
7  & oop C-H wag & 108.67 & 121.95 & 109.20 \\ \midrule
8  & oop C-H wag & 111.53 & 132.82 & 116.80 \\ \midrule
9  & oop C-H wag & 121.07 & 134.73 & 122.90 \\ \midrule
10 & oop C-H wag & 129.00 & 139.51 & 122.00 \\ \midrule
11 & ring stretch & 133.03 & 138.52 & 127.70 \\ \midrule
12 & ring stretch & 133.22 & 139.74 & 124.85 \\ \midrule
13 & ring stretch + C-H bend & 135.25 & 142.19 & 132.42 \\ \midrule
14 & ring stretch + C-H bend  & 136.28 & 145.24 & 132.91 \\ \midrule
15 & ring stretch + C-H bend  & 164.30 & 144.32 & 142.09 \\ \midrule
16 & ring stretch + C-H bend & 167.09 & 165.37 & 150.76 \\ \midrule
17 & ring stretch + C-H bend & 175.17 & 160.81 & 152.13 \\ \midrule
18 & ring stretch + C-H bend & 181.30 & 185.15 & 168.00 \\ \midrule
19 & ring stretch & 188.57 & 197.17 & 178.17 \\ \midrule
20 & ring stretch + C-H bend & 197.44 & 203.59 & 183.74 \\ \midrule
21 & ring stretch + C-H bend & 213.30 & 219.70 & 195.15 \\ \midrule
22 & ring stretch + C-H bend & 227.51 & 221.27 & 196.02 \\ \midrule
23 & C-H stretch & 236.81 & 411.44 & 378.40 \\ \midrule
24 & C-H stretch & 360.20 & 411.48 & 375.05 \\ \midrule
25 & C-H stretch & 365.61 & 413.00 & 381.72 \\ \midrule
26 & C-H stretch & 381.04 & 415.31 & 376.04 \\ \midrule
27 & C-H stretch & 386.05 & 416.34 & 383.11 \\
\bottomrule
\end{longtable}

\nocitesup{*}
\bibliographystylesup{apsrev4-1}
\bibliographysup{Supp}

\end{document}